# Direct current voltage induced by microwave signal in a ferromagnetic wire

A.Yamaguchi[1], K. Motoi[1], H. Miyajima[1], and Y. Nakatani[2]

[1]Department of Physics, Keio University, Hiyoshi 3-14-1, Yokohama, Kanagawa 223-8522, Japan

[2]University of Electro-communications, Chofugaoka 1-5-1, Chofu, Tokyo 182-8585, Japan

**[ABSTRACT]**

Experimental results of rectification of a constant wave radio frequency (RF) current flowing in a single-layered ferromagnetic wire are presented. We show that a detailed external magnetic field dependence of the RF current induced a direct-current voltage spectrum. The mechanism of the rectification is discussed in a term of the spin transfer torque, and the rectification is closely related to resonant spin wave excitation with the assistant of the spin-polarized RF current. The micromagnetic simulation taking into account the spin transfer torque provides strong evidence which supports the generation of spin wave excitation by the RF current.

# I. INTRODUCTION

Recently it is found that the spin and electric charge of electrons in nano-artificial magnetic systems exhibit peculiar behaviors in the radio-frequency (RF) region through ferromagnetic resonance (FMR) which has led to the spin-polarized current in the RF region. Furthermore, spin wave excitation and magnetization reversal is induced by the spin-polarized RF current. From the view points of fundamental nano-scale magnetism and technical spintronics, many theoretical [1, 2] and experimental [3 - 16] studies have been progressed and focused on a system with spatial distribution of magnetization such as magnetic multilayers [3 – 10], magnetic tunnel junctions (MTJ) [11], spin-valves [12], and single-layered ferromagnetic wires [13 – 15]. These systems have numerous possible technical applications in magnetoresistive random access memory (MRAM), microwave generators [9, 10], and so on. The physical mechanism, as revealed by the experimental results, is described in term of the spin angular momentum transferring which occurs due to the interaction between the spin-polarized current and the magnetic moment. One of the most interesting properties is the rectification of RF current. When the resonant RF current flows through a system with magnetization distribution, it resonantly excites spin wave in the system and the magnetic moment precesses around the effective magnetic field consisting of magnetic anisotropy field, demagnetizing field and external field. As a result, the direct-current (DC) voltage is produced by the magnetoresistance oscillation due to FMR generated by the spin-transfer effect or RF field [11 - 19].

In the previous studies of pillar structure systems consisting of two ferromagnetic layers (a spin-polarizer layer and a free layer) have been adopted to originate the spin-transfer effect. On the other hand, it has been clarified that the spin dynamics is excited by the spin-transfer torque due to the spin-polarized current flowing through a system with domain wall or magnetic vortex core [13 - 16].

In this paper, we present detail experimental results of the rectifying effect observed in a single-layered ferromagnetic nanowire [13]. As mentioned above, the rectifying effect is attributable to the resonant excitation of spin waves by the RF current. $I(t) = I_0 \cos \omega_1 t$ at time $t$. Once the spin wave is excited in the wire, the magnetic moment precesses around the effective magnetic field with the angular frequency $\omega_1$, the resistance change $\Delta R(t)$ will oscillate as a function of $\cos \omega_2 t$, originating from the anisotropic magnetoresistance (AMR) effect. The DC voltage given by $V(t) = I(t) \cdot \Delta R(t)$ is generated by a term involving mixing between RF current and magnetoresistance;

$$V(t) = V_0 \left[ \cos(\omega_1 - \omega_2)t + \cos(\omega_1 + \omega_2)t \right]/2 \qquad (1)$$

If $\omega_1 = \omega_2 = \omega$, $V(t) = V_0 (1 + \cos 2\omega t)/2$; that is, the nanowire acts as a detector which turns the RF signal to DC voltage with second harmonic oscillation.

## II. RECTIFYING EFFECT OF RF CURRENT

We focus on a single layered $Ni_{81}Fe_{19}$ wires [Fig. 1]. The wires are fabricated on MgO substrates using the procedure described in Ref. 13. We examine 30 and 50-nm-thick $Ni_{81}Fe_{19}$ nanowires and a 20-nm-thick Au nanowire contacted with Au electrode, of the 100 nm thick, as shown in Fig. 1. The width of the 50-nm-thick $Ni_{81}Fe_{19}$ nanowires is 300, 650, and 2200 nm and that of 30-nm-thick $Ni_{81}Fe_{19}$ and Au nanowires is 5 μm and 300 nm, respectively.

A ground-signal-ground (GSG) type microwave prove is connected to the nanowire, and the bias-tee circuit detects the DC voltage difference induced by the RF current flowing through the nanowire, as schematically illustrated in Fig. 1. The sinusoidal constant wave RF current is injected into the nanowire by a signal generator with a frequency range from 10 MHz to 40 GHz. The external magnetic field $\mathbf{H}_{ext}$ is applied in the substrate plane as a function of angle $\theta$ to the longitudinal axis of the nanowire, while the Oersted field produced by the RF currents flowing in the loop of the Au electrodes generates an additional field at the sample. However, the field is canceled out at the nanowire position because of the symmetrical arrangement of the Au electrode. All experiments are performed at room temperature.

Figure 5 shows the magnetic field dependence of the DC voltage difference generated in the wires with widths of 5 μm. The RF current density injected into the wire is about $2.3 \times 10^{10}$ A/m$^2$, the value of which is relatively low for the Joule heating to be applied. According to our recent estimation of the current density from the measurement of the RF characteristic impedance of the wire, the density is found to be almost same to the value. An external magnetic field is applied in the substrate plane at the angle of (a) $0°$, (b) $30°$, (c) $60°$, and (d) $90°$ to the wire-axis. As seen in the figure, the peak position of resonance spectrum shifts to the higher frequency region with the external magnetic field. The sign of the DC voltage reverses when the angle of the applied magnetic field is changed form $\theta$ to $\theta + 180°$. On the other hand, no similar effects are observed in the 20-nm-thick Au nanowires, which strongly indicates that the resonance is of magnetic origin and generated by the RF current.

Figure 3 shows the magnetic field dependence of the resonance frequency of the nanowire with widths of 300, 650, 2200, and 5000 nm, where the magnetic field is applied at the angle of $45°$ to the wire-axis. The resonance frequency increases with increasing magnetic field but decreasing wire-width, which means that the resonance mode possibly reflects the shape magnetic anisotropy.

According to the extended Kittel's equation [20], the ferromagnetic resonance frequency at the magnetic field $H$ is approximately given by

$$\omega(H) = \frac{g\mu_B\mu_0}{\hbar} \cdot \left[(H + H_A + M_S) \cdot (H + H_A)\right]^{\frac{1}{2}}, \qquad (2)$$

where $M_S$ denotes the saturation magnetization, $g$ the Lande factor, $\hbar$ Planck's constant, $\mu_0$ the permeability of free space, $\mu_B$ the Bohr magneton, and $H_A$ the effective anisotropy field including demagnetizing and anisotropy fields in the substrate plane. The experimentally determined $H_A$ for the nanowires with widths of 300, 650, 2200, and 5000 nm is 1160, 830, 260, and 100 Oe, respectively. The resonance frequency is in excellent agreement with that of calculation using the extended Kittel's eq.(2), indicating the existence of resonant spin wave excitation. As the RF current induces the precession of the magnetic moments in the nanowires and the excitation of the spin wave, the resistance due to the AMR effect oscillates at the resonance frequency. This resistance oscillation generates the DC voltage by combining with the RF current, as observed in the case of the spin-torque diode effect in the pillar structure systems [11, 12].

## III. MICROMAGNETICS CALCULATION

The current-induced dynamics of the magnetic moment are analytically calculated by the micromagnetic simulations based on the Landau-Lifshitz-Gilbert (LLG) equation including the spin-transfer term [16, 21]. The modified LLG equation is given by

$$\frac{\partial \mathbf{m}}{\partial t} = -\gamma_0 \mathbf{m} \times \mathbf{H}_{eff} + \alpha \mathbf{m} \times \frac{\partial \mathbf{m}}{\partial t} - (\mathbf{u}_s \cdot \nabla) \mathbf{m}, \qquad (3)$$

where $\mathbf{m}$ denotes the unit vector along the local magnetization, $\gamma_0$ the gyromagnetic ratio, $\mathbf{H}_{eff}$ the effective magnetic field including the exchange and demagnetizing fields, and $\alpha$ the Gilbert damping constant, respectively. The last term in eq. (3) represents the spin-transfer torque which describes the spin transferred from conduction electrons to localized spins. The spin-transfer effect is a combined effect of the spatial non-uniformity of magnetization distribution and the current flowing. The vector with dimension of velocity, $\mathbf{u}_s = -\mathbf{j} P g \mu_B / (2eM_s)$, is essentially the spin current associated with the electric current, where $\mathbf{j}$ is the current density, $P$ the spin polarization of the current, and $e$ the electronic charge, respectively. The spatial non-uniformity of the magnetization distribution in a nanowire is mainly caused by the spatial dispersion of the demagnetizing field.

A micromagnetic simulation of the system considering the spin transfer torque is performed to confirm the generation of spin wave excitation by the constant wave RF current [16, 21]. The nanowire adopted in the simulation has a width and length of 300 nm and 4 μm, respectively. The parameters used for the calculation are; the unit cell size of 4 nm × 4 nm with constant thickness 50 nm, the magnetization 1.08 T, the spin current density $u$ = 3.25 m/s, and the spin polarization $p$ = 0.7 [16]. In the simulation, the external magnetic field of 200 Oe is applied in the substrate plane at the angle of $45°$ from the wire-axis, where the magnitude of the field is not large enough to direct perfectly the magnetization along the direction of the applied magnetic field. The local spins in the nanowire are almost parallel to the effective anisotropy field.

Figure 4 (a) shows a snapshot of the map and the cross-section of the z component of the magnetization in the middle of the simulation. The black and white stripes correspond to the –z and +z components of the magnetization, respectively, and the stripe magnetization pattern due to the spin wave excitation induced by the RF current is clearly observed. It should be noted that the period of the stripe pattern of approximately 100 nm is consistent with the spin wave length estimated by $\sqrt{A/K}$, where $A$ and $K$ are the exchange stiffness constant and the anisotropy energy written by $K = g\mu_B\mu_0 H_C$, respectively. The magnetization response is simulated as a function of the RF current frequency. The mean values of the y and z components of magnetization are shown in Fig. 4 (b) as a function of the RF frequency. The mean value of x component is not suitable for the analysis as it has large background because of the uniaxial magnetic anisotropy. As shown in Fig. 4 (b), both y and z components peak at 10.4 GHz, which is consistent with the spin wave resonance frequency. The simulation result is in accord with the experimental result.

**IV. SIMPLE ANYLYTICAL MODEL AND EXPERIMENTAL RESULT**

Here, we derive an analytical expression for the DC voltage produced by the spin wave excitation. Suppose we have an ($x$, $y$, $z$) coordinate system: each component corresponds to the short wire axis, longitudinal wire axis, and normal to the substrate plane, respectively, as schematically shown in Fig. 5 (a). The external magnetic field $\mathbf{H}_{ext}$ is applied along the angle $\theta$ from the y-coordinate axis. Then, let us define a new coordinate system ($a$, $b$, $c$) where the $b$ direction corresponds to the equilibrium direction of unit magnetization $\mathbf{m}_0$ along an effective magnetic field $\mathbf{H}_{eff}$ including $\mathbf{H}_{ext}$ and $\mathbf{H}_A$ (the shape magnetic anisotropy field). The unit vector $\mathbf{m}_0$ inclines at an angle $\psi$ to the y-coordinate axis, and $\mathbf{H}_A$ is parallel to the y-coordinate axis. The magnetization precession around $\mathbf{H}_{eff}$ only results in a small time-dependent component of the magnetization perpendicular to $\mathbf{m}_0$. The assumption is valid whenever the spin wave excitation is small. Then, we decompose the unit vector $m$ in the ($a$, $b$, $c$) coordinate system as $\mathbf{m}(t) = \mathbf{m}_0 + \delta\mathbf{m}(t)$, considering a small variation $\delta\mathbf{m} = (m_a(t), 0, m_c(t))$ and $\mathbf{m}_0 = (0,1,0)$. The vector $m$ is obtained from eq.(3):

$$\frac{\partial \mathbf{m}}{\partial t} \approx -\gamma_0 \times \mathbf{H}_{eff} + \alpha \mathbf{m}_0 \times \frac{\partial \delta\mathbf{m}}{\partial t} - \mathbf{s}, \tag{4}$$

where

$$\mathbf{m} = (m_a, m_b, m_c), \quad m_b = \sqrt{1-(m_a^2 + m_c^2)}, \tag{5}$$

and the spin transfer torque $\mathbf{S}$ almost perpendicular to $\mathbf{m}_0$ in the ($a$, $b$, $c$) coordinate system is defined as

$$\mathbf{s} = (\mathbf{u}_s \cdot \nabla)\mathbf{m} = j \cdot (s_a \cos\psi, -s_b \sin\psi, s_c), \tag{6}$$

The effective field $\mathbf{H}_{eff}$ is expressed in the ($a$, $b$, $c$) coordinate system as

$$\mathbf{H}_{eff} = \begin{pmatrix} H_a \\ H_b \\ 0 \end{pmatrix} = \begin{pmatrix} H_{ext}\sin(\theta-\psi) - H_A \sin\psi \\ H_{ext}\cos(\theta-\psi) + H_A \cos\psi \\ 0 \end{pmatrix}. \tag{7}$$

Substituting eq. (5), (6), and (7) into eq. (3), we obtain the magnetization dynamics;

$$\frac{\partial}{\partial t}\begin{pmatrix} m_a \\ m_b \\ m_c \end{pmatrix} \approx -\gamma_0 \begin{pmatrix} -m_c H_b \\ m_c H_a \\ m_a H_b - m_b H_a \end{pmatrix} + \alpha \frac{\partial}{\partial t}\begin{pmatrix} m_c \\ 0 \\ -m_a \end{pmatrix} - j\begin{pmatrix} s_a \cos\psi \\ -s_b \sin\psi \\ s_c \end{pmatrix} \tag{8}$$

Under the conditions $|\mathbf{m}|=1$ and $m_b \approx 1$, eq. (8) is reduced to,

$$\frac{\partial}{\partial t}\begin{pmatrix} m_a \\ m_c \end{pmatrix} \approx -\gamma_0 \begin{pmatrix} -m_c H_b \\ m_a H_b - H_a \end{pmatrix} + \alpha \frac{\partial}{\partial t}\begin{pmatrix} m_c \\ -m_a \end{pmatrix} - j\begin{pmatrix} s_a \cos\psi \\ s_c \end{pmatrix}, \tag{9}$$

assuming that the time-dependent magnetization precession angle around the direction of $\mathbf{m}_0$ is parallel to $\mathbf{H}_{eff}$ and that the angle is very small. The solution of eq. (9) can be presented by the Fourier transformation;

$$\begin{pmatrix} m_a \\ m_c \end{pmatrix} = \frac{1}{\omega^2 - (\omega_k + i\omega\alpha)^2} \begin{pmatrix} i\omega & \omega_k + i\omega\alpha \\ -(\omega_k + i\omega\alpha) & i\omega \end{pmatrix} \begin{pmatrix} js_a \cos\psi \\ js_c - \gamma_0 H_a \end{pmatrix}, \quad (10)$$

where $\omega_k = \gamma_0 H_b$ denotes the spin wave resonant angular frequency corresponding to $\omega(H)$ given by eq.(2).

Due to the spin wave excitation with precession angle of $\phi_l$, the small variation in the magnetic moment around the direction of $\mathbf{H}_{eff}$ induces the time-dependent AMR effect. When $\cos^2(\phi_l) \approx 1$ and $\sin^2(\phi_l) \approx 0$ for small $\phi_l$, we can approximately derive the average time-dependent resistance $\langle R(t) \rangle$ in each cell;

$$\langle R(t) \rangle = \left\langle \sum_\ell \Delta R \cos^2(\psi + \phi_\ell(t)) \right\rangle \approx \Delta R \left[ \cos^2\psi - \frac{1}{2} \langle \sin 2\psi \sin 2\phi_\ell(t) \rangle \right], \quad (11)$$

where $\Delta R$ denotes the resistance change due to the AMR effect, the number $l$ the index of the domain of the spin wave excitation, and $\sin 2\phi_\ell \approx 2(\Delta m/|m|)_a$ by using the $a$ component of $\delta\mathbf{m}$. When the RF current given by $I(t) = I_0 \cos(\omega t)$ flows in the wire, the frequency spectrum of the induced voltage is expressed by the Fourier transformation of $V(t) = R(t)I(t)$;

$$V(\omega) = A(\omega) I_0^2 \sin 2\psi \cos\psi, \quad (12)$$

where $A(\omega)$ is written by

$$A(\omega) \approx \frac{2\omega^2 \alpha \omega_k \cdot s_a + (\omega^2 - \omega_k^2) \omega_k \cdot (s_c - \gamma_0 H_a / j) \cos\psi^{-1}}{(\omega^2 - \omega_k^2)^2 + 4\omega^2 \alpha^2 \omega_k^2} \cdot \frac{\Delta R}{b \cdot d}, \quad (13)$$

where $b$ and $d$ are the width and thickness of the nanowire, respectively.

In the case that $|\mathbf{H}_{ext}|$ is higher than $|\mathbf{H}_A|$, the magnetic moment directs almost along $\mathbf{H}_{ext} \approx \mathbf{H}_{eff}$. Then, we obtain $\psi \approx \theta$ ( see Fig. 5 (a)), which well corresponds to the result described in Ref 13. The relation $\psi \approx \theta$ gives the frequency variation of the induced voltage $V(\omega)$ proportional to $I_0^2 \sin 2\theta \cos\theta$. On the contrary, $\mathbf{H}_{ext}$ is smaller than $\mathbf{H}_A$, the magnetic moment is almost parallel to the direction of $\mathbf{H}_{eff}$. Then, $0° < \psi = \text{const.} < \theta$, and the $V(\omega)$ is almost proportional to $H_{ext} \sin\theta$. This means that the out-of-plane component of magnetization leads to the dispersion shape.

As a result, the field and angle dependences of the induced DC voltage $V(\omega)$ are summarized as follows:

(i)     $V(\omega) \propto \sin 2\theta \cos\theta$ in $|\mathbf{H}_{ext}| \gg |\mathbf{H}_A|$,

(ii)    $V(\omega) \propto H_{ext} \sin\theta$ in $|\mathbf{H}_{ext}| < |\mathbf{H}_A|$

Figures 5 (b), (c), and (d) show the angle dependence of the induced DC voltage at the resonance frequency in the magnetic field of 400, 100, and 30 Oe, respectively. The inset of Fig.5 (b) shows the experimental result of the magnetoresistance for 30 nm-thick $Ni_{81}Fe_{19}$ wire with 5 μm width in the magnetic field perpendicular to the longitudinal wire axis ($\theta = 90°$).

The result of the magnetoresistance measurement in the inset of Fig. 5 (b) offers evidence that the magnetization is almost parallel to the direction of the field $\mathbf{H}_{ext}$ at $|\mathbf{H}_{ext}| = 400$ Oe. Then, according to the analytical model described above, the angle dependence of the induced DC voltage $V(\omega)$ should be proportional to $\sin 2\theta \cos\theta$, corresponding to the case (i). In Fig. 5 (b), the dashed-line represents the $\sin 2\theta \cos\theta$ curve, and it is in good agreement with the experimental results. The magnetization is almost along the direction of $\mathbf{H}_A$ at $|\mathbf{H}_{ext}| = 30$ Oe. Then, the angle dependence of the induced DC voltage $V(\omega)$ corresponds to the case (ii), that is $V(\omega) \propto H_{ext} \sin\theta$. In Fig. 5 (d), the experimental result is consistent with the $\sin\theta$ curve. As shown in Figs. 5 (b) and (d), we found that the simple analytical model expression given by eqs. (12) and (13) qualitatively fits well to the experimental results. The angle dependence of $V(\omega)$ at $|\mathbf{H}_{ext}| = 100$ Oe is intermediate with the angle dependence between the cases (i) and (ii) as shown in Fig. 5 (c).

The spectra shape is determined by $A(\omega)$ of eq. (13). $A(\omega)$ is given by the superposition of the Lozentzian function with a maximum at $\omega = \omega_k$ and the dispersion function. The component of the Lozentzian function basically governs $A(\omega)$, indicating that the in-plane spin transfer component determines the spectrum shape. However, as the angle $\theta$ increases, the component of the dispersion function in $A(\omega)$ increases. This is consistent with the actual spectra shapes observed in the experimental results shown in Fig. 2. The additional wiggles in the trace of Fig. 2 correspond to the additional spin wave modes derived from the complicated spin wave excitation. While, if the in-plane RF field induce the magnetization dynamics, the spectrum shape is described only the dispersion function [19].

The solution of eqs. (12) and (13) can be also represented as follows:

$$V(\omega) = I \cdot \Delta R \cdot \Gamma(\omega) \cdot \sin 2\psi \cos \psi, \tag{14}$$

$$\Gamma(\omega) \approx \frac{2\omega^2 \omega_k \alpha \cdot js_a + \omega_k (\omega_k^2 - \omega^2) \cdot (js_c - \gamma_0 H_a) \cos \psi^{-1}}{(\omega_k^2 - \omega^2)^2 + 4\alpha^2 \omega^2 \omega_k^2}. \tag{15}$$

Here, $\Gamma(\omega)$ gives the dimensionless value and the spectrum shape. We can fit the change in the normalized DC voltage $V(\omega)/I\Delta R$ observed using the eq.(14). With the in-plane and the out-of-plane component of the spin transfer torque $js_a$, $js_c$, and the Gilbert damping parameter $\alpha$ as fit parameters, the typical signals measured in the wire with width of 5 μm in the application of the external field $|\mathbf{H}_{ext}| = 400$ Oe could be fitted as shown in Fig. 6. These fits allow us to determine those parameters, which were found to be $(js_a, js_c, \alpha) = (4.45 \times 10^6, 6.79 \times 10^5, 0.016)$ at $\theta = 15°$ and $(js_a, js_c, \alpha) = (1.45 \times 10^6, 3.13 \times 10^6, 0.010)$ at $\theta = 30°$ in the wire with width of 5 μm. The Gilbert damping constants determined by the fittings agree with the values reported by Costache *et al*. [18], and they are larger than the value commonly accepted for a thin film of $Ni_{81}Fe_{19}$ [22]. In addition, we note here that the DC voltage spectrum shape, comprising of the superposition of the Lorentzian function and the dispersion function, changes to a more complex and wiggle shape as the frequency and angle applied external field change. These can be due to the spatial inhomogeneities of the incoherent spin wave excitation and out-of-plane component neglected in the time-averaging resistance.

Electrical conduction in a ferromagnetic metal generally depends on the direction of the magnetization. The phenomenological description of electrical field **E** given by the two current model takes the following form

$$\mathbf{E} = \rho_\perp \mathbf{j} + \mathbf{m}(\mathbf{j} \cdot \mathbf{m}) \cdot (\rho_\parallel - \rho_\perp) + \rho_{H_0} \mathbf{m} \times \mathbf{j}. \qquad (16)$$

Here, **j** denotes the electrical current, **m** denotes the unit vector along the local magnetization, $\rho_\perp$, $\rho_\parallel$ the resistivity components refer to the parallel and perpendicular directions of **j** with respect to **m**, $\rho_{H_0}$ the extraordinary Hall resistivity component, respectively. Our result derived from the analytical model given by eq. (3) is consistent with that of the eq. (16) extended by Juretschke [17]. Juretschke introduced the phenomenological oscillating component of the magnetization to eq. (16) and obtained the generation of the DC voltage, assuming the small angle of magnetization precession caused by the RF field or something [17]. Here, what is the origin of the magnetization dynamics? The RF field in the substrate plane and perpendicular to the plane causes the magnetization precession. In the former case, the angle dependence of the induced DC voltage agrees with the case (i) or (ii). However, in the later case, it is proportional to $\sin 2\theta$ [19]. In our case of this study, the RF field is canceled out at the wire position due to the symmetrical arrangement of the electrodes.

Instead, the spin-polarized RF current flowing through the ferromagnetic wire, with the spatial non-uniformity of the magnetization distribution caused by the spatial dispersion of the demagnetizing field, produces the spin transfer torque, and it generates the magnetization dynamics.

The DC voltage originates from the magnetoresistance oscillation due to the magnetization precession generated by the spin-transfer effect at the FMR frequency. We measure the DC voltage induced by RF signal along the major and minor axes of the single-layered ferromagnetic wire. The $DC_L$ and $DC_S$ voltage, which respectively corresponds to that induced along the major-axis and minor-axis of the wire, is shown in Fig. 7 in the case of $H$ = 500 Oe and $\theta = 30°$, where the solid and dashed lines indicate the $DC_L$ and $DC_S$ voltage spectra, respectively. The rectifying effect is interpreted in terms of the magnetoresistance oscillation and the planer Hall effect, as introduced by eq. (16).

## V. CONCLUSION

(1) The resistance oscillation due to the spin wave excitation caused by spin-polarized RF current generates the DC voltage by combing with the RF current in the single layered ferromagnetic wires.

(2) The experimental results for the rectification of RF current flowing in the wire in the

presence of an external magnetic field show the existence of a regime: with the precession of the average magnetization around the effective anisotropy magnetic field including the external and demagnetizing magnetic field.

(3) The applied magnetic field angle $\theta$ dependence of the induced DC voltage is proportional to $\sin 2\theta \cos \theta$ in the high magnetic field and $\sin \theta$ in the low filed, respectively.

The rectifying effect of the RF current results in a highly-sensitive measurement of spin dynamics in nano-scale magnets. This effect facilitates the development of spintronics devices such as radio-frequency detectors in telecommunication [23], magnetic sensors, and coplanar-type high-frequency band-pass filters.

**ACKNOWLEDGMENTS**

The present work was partly supported by a Grants-in-Aid from the Ministry of Education, Culture, Sports, Science, and Technology, Japan, by the Keio Leading-edge Laboratory of Science and Technology project 2007, and by the Foundation Advanced Technology Institute, Japan.

**Figure caption**

Figure 1

The radio frequency (RF) electric circuit and the optical micrograph (top view) of the system consisting of electrodes and the magnetic nanowire. The magnetic field is applied in the substrate plane at an angle of $\theta$ to the longitudinal axis of the nanowire. The direct-current (DC) voltage across the wire is measured by using a bias tee.

Figure 2

The DC voltage response generated in the 30 nm-thick $Ni_{81}Fe_{19}$ wire with 5 μm width in response to the RF current as a function of RF current frequency when the external magnetic field is applied at an angle of (a) $\theta = 0°$, (b) $\theta = 30°$, (c) $\theta = 60°$, and (d) $\theta = 90°$. Each DC response is vertically shifted for clarity.

Figure 3

Resonance frequency as a function of the magnetic field. The circle, square, up-triangle, and down-triangle marks denote the experimental data points for the nanowires with widths of 300, 650, 2200, and 5000 nm, respectively. The solid-line curve represents the best fitting to the extended Kittel's equation.

Figure 4

(a) Typical micromagnetic simulation including the spin transfer and field-like torques (shown in inset). The simulated pattern (black and white stripes) shows the oscillating magnetization in a magnetic nanowire. The black stripe corresponds to the –z component of the magnetization; the white stripe, the +z component. (b) The y and z components of the magnetic moment are shown as functions of the RF current frequency. The red and black lines correspond to the y and z components, respectively.

Figure 5

(a) Schematic coordinate system adopted in the simple analytical calculation model. The longitudinal wire axis is parallel to the y-axis. Angle dependence $\theta$ of the DC voltage amplitude when the external magnetic field of (b) 400 Oe, (c) 100 Oe, and (d) 30 Oe was applied in the substrate plane, respectiely. The inset shows the result of magnetoresistance measurement when the field was applied at $\theta = 90°$. The dashed-line in (b) and (c) represents a $\sin 2\theta \cos \theta$ curve, and the dashed-line in (d) corresponds to a $\sin \theta$ curve.

Figure 6
Measured DC voltage as a function of the RF current frequency in the wire with width of 5 μm. The solid lines are the fit to the data using eq. (15). The dotted and dashed lines correspond to the in-plane and out-of-plane component, respectively.

Figure 7

The $DC_L$ and $DC_S$ voltage as a function of RF current frequency in $H$= 500 Oe at $\theta$ = 30°.

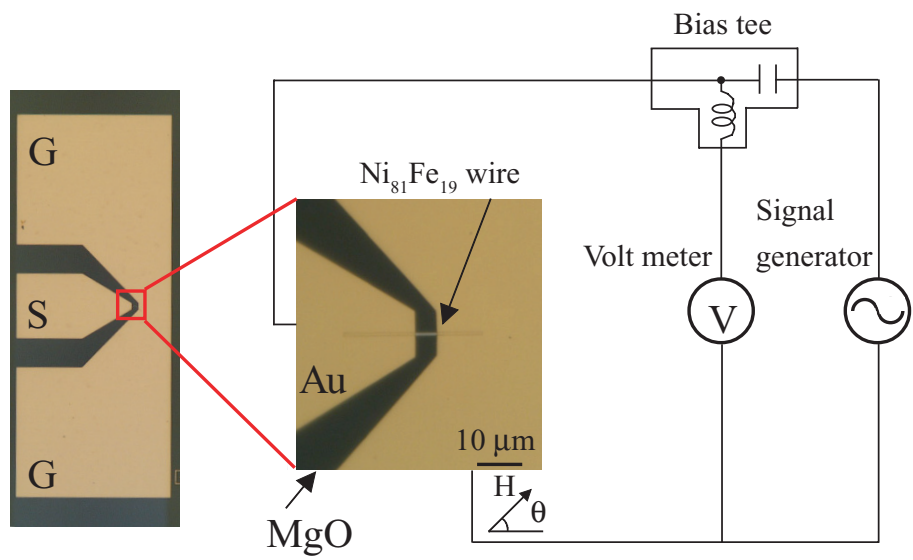

Fig. 1 A. Yamaguchi *et al*.

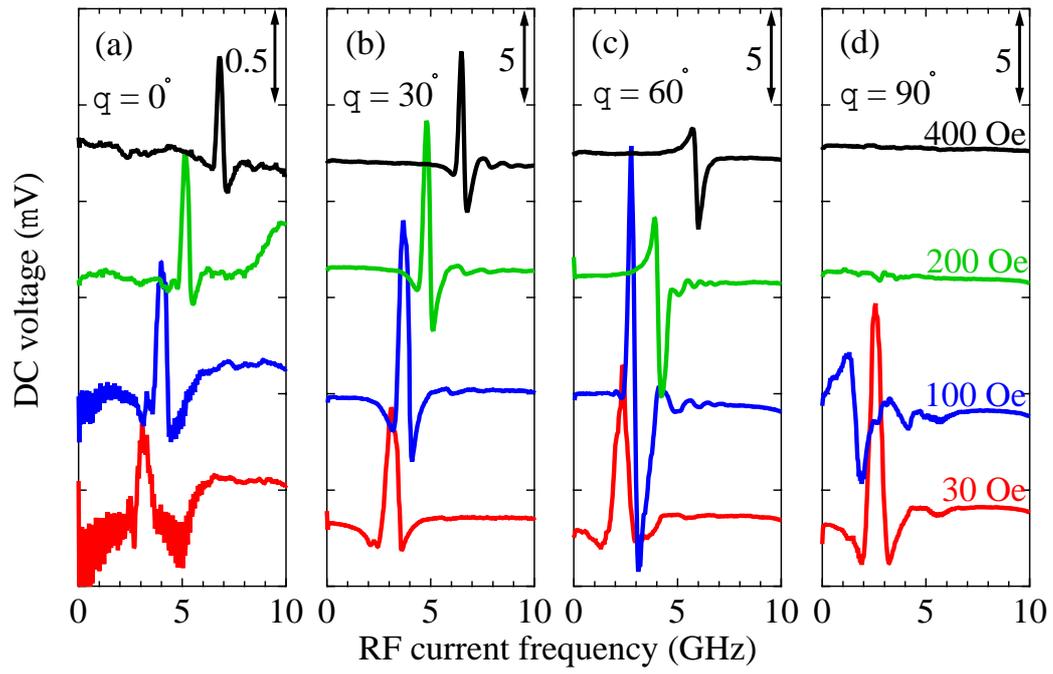

Fig. 2 A. Yamaguchi *et al*.

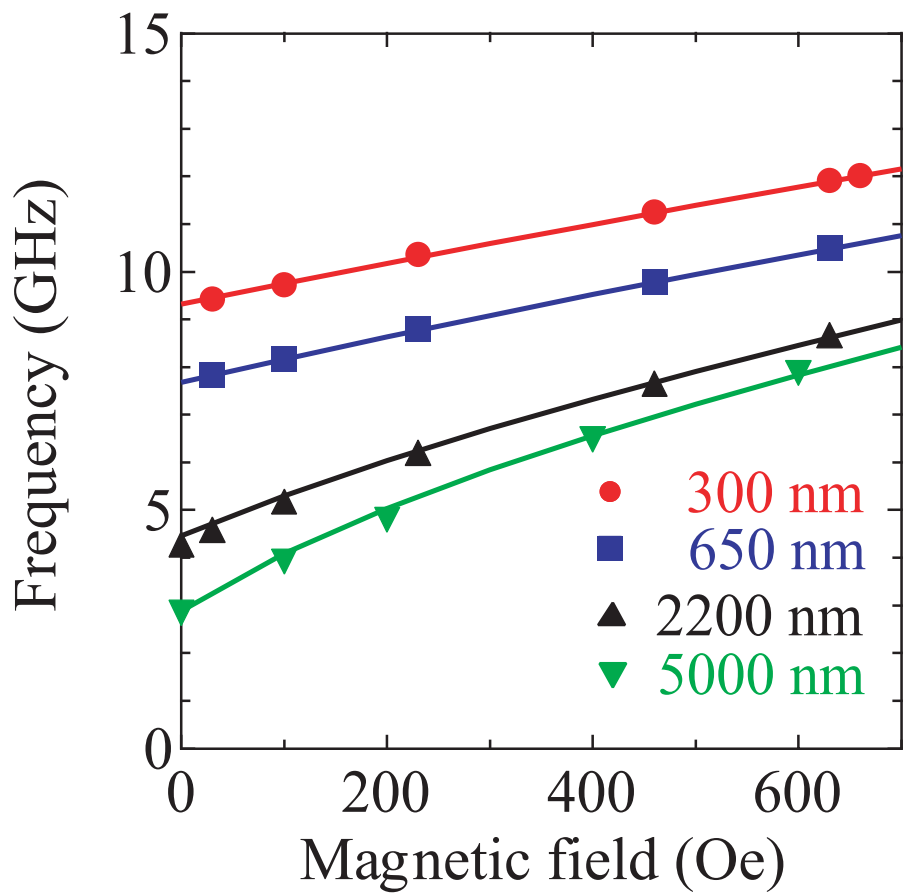

Fig. 3 A. Yamaguchi *et al*.

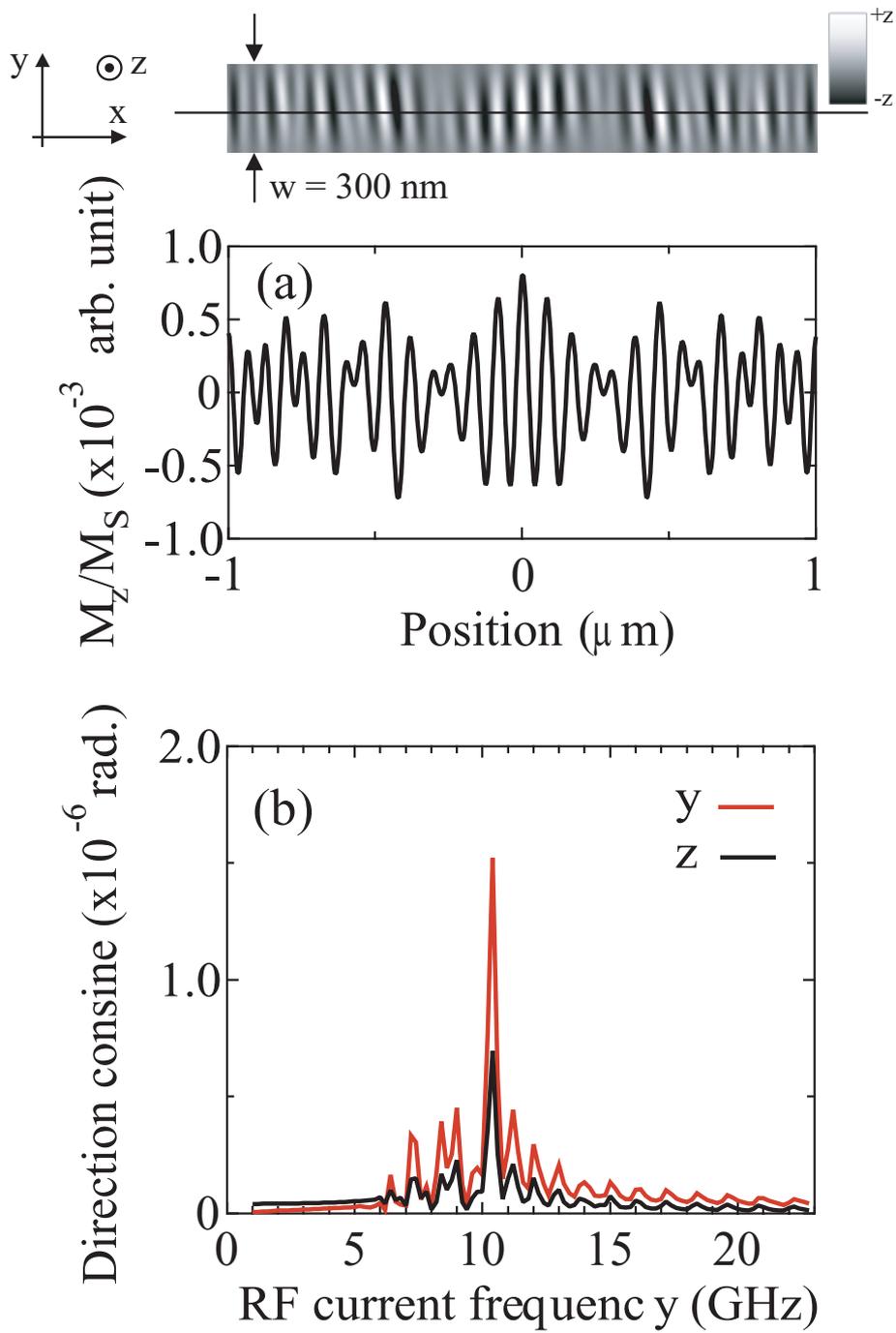

Fig. 4 A. Yamaguchi *et al.*

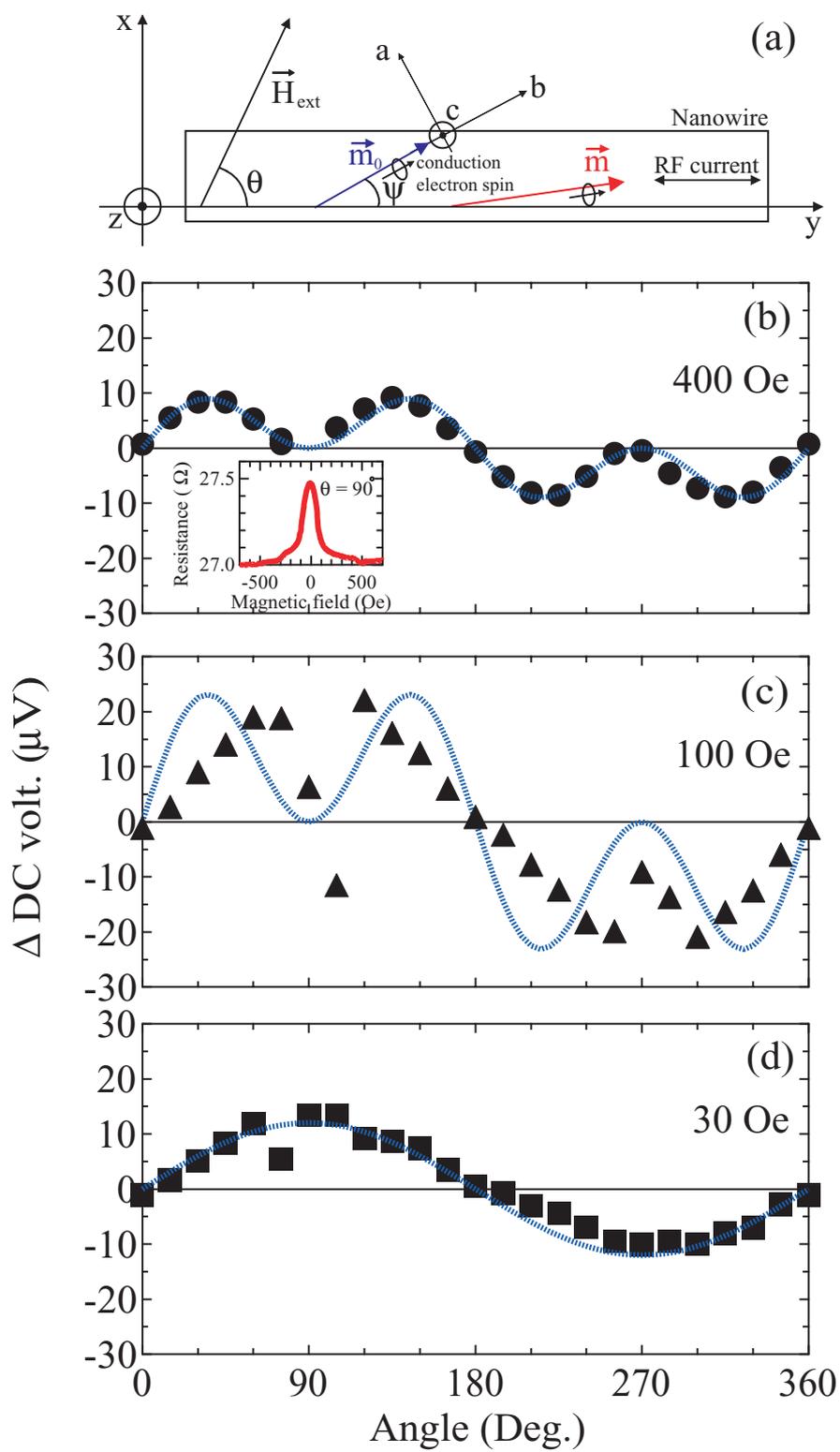

Fig. 5 A. Yamaguchi *et al*.

27/27

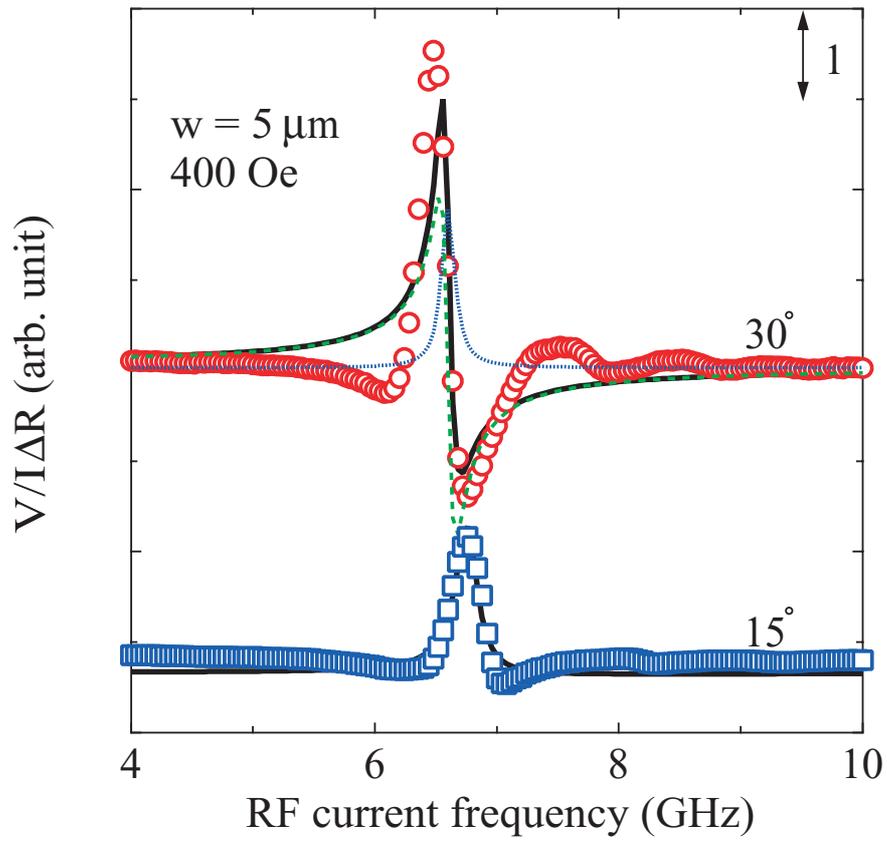

Fig. 6 A. Yamaguchi *et al*.

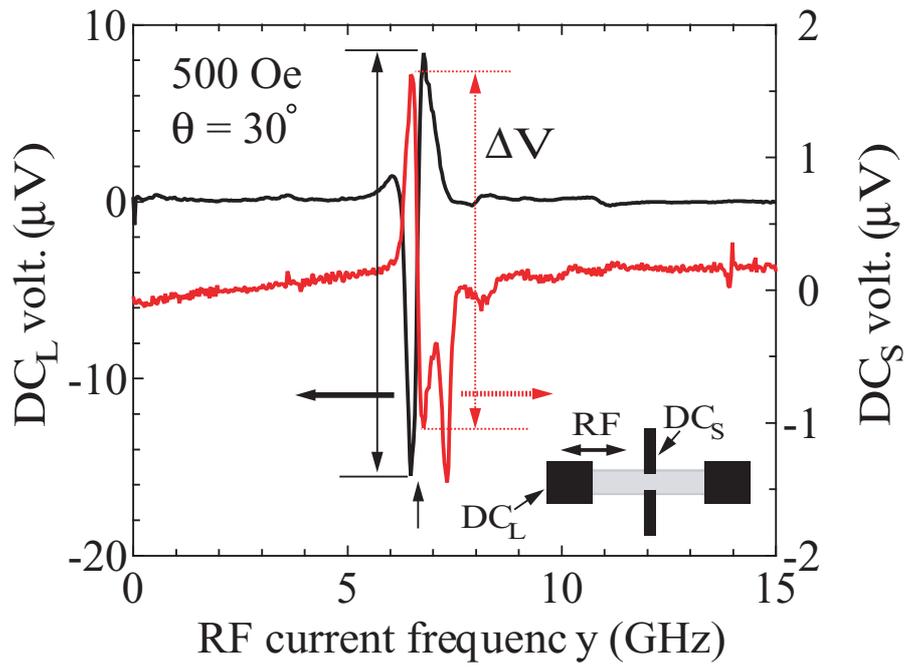

Fig. 7 A. Yamaguchi *et al*.